\documentclass[runningheads]{llncs}
\usepackage{bdlexp}
\usepackage[hidelinks]{hyperref}
\hypersetup{colorlinks=true,linkcolor=blue,citecolor=blue,urlcolor=blue}
\usepackage{orcidlink}
\usepackage{version}

\newcommand{\doi}[1]{{doi:\href{https://doi.org/#1}{\nolinkurl{#1}}}}
\renewcommand{\url}[1]{{\href{#1}{\nolinkurl{#1}}}}

\spnewtheorem{sdef}{Definition}{\bfseries}{\rmfamily}

\title{The Interdefinability of Expansions of Belnap-Dunn Logic}
\author{C.A. Middelburg\,\orcidlink{0000-0002-8725-0197}}
\institute
 {Informatics Institute, Faculty of Science, University of Amsterdam \\
  Science Park~900, 1098~XH Amsterdam, the Netherlands \\
  \email{C.A.Middelburg@uva.nl}}

\titlerunning
 {The Interdefinability of Expansions of Belnap-Dunn Logic}
\authorrunning
 {C.A. Middelburg}

\begin{document}

\maketitle

\begin{abstract}
Belnap-Dunn logic, also knows as the logic of First-Degree Entailment, 
is a logic that can serve as the underlying logic of theories that are 
inconsistent or incomplete.
For various reasons, different expansions of Belnap-Dunn logic with 
non-classical connectives have been studied.
This paper investigates the question whether those expansions are 
interdefinable with an expansion whose connectives include only 
classical connectives.
Surprisingly, this relevant question is not addressed anywhere in the 
published studies.
The notion of interdefinability of logics used is based on a general 
notion of definability of a connective in a logic that seems to have 
been forgotten.
Attention is also paid to the extent to which the expansion whose 
connectives include only classical connectives is related to the 
version of classical logic with the same connectives.
\begin{msc-classcode} 
03B50 (Primary)\,\, 03B53 (Secondary) 
\end{msc-classcode}
\end{abstract}

\section{Introduction}
\label{INTRO}

The main aim of this paper is to gain insight into the interdefinability 
of expansions of Belnap-Dunn logic (\BDL)~\cite{Bel77a,Bel77b}.
Belnap-Dunn logic, also knows as the logic of First-Degree 
Entailment, is a logic that can serve as the underlying logic of 
theories that are inconsistent or incomplete.
Interestingly, the logical consequence relation of \BDL\ is included in 
the logical consequence relations of three well-known logics, to wit 
Priest's Logic of Paradox (\LP)~\cite{Pri79a}, Kleene's strong 
$3$-valued logic (\Kiii)~\cite{Kle52a}, and the version of classical 
propositional logic with the same connectives as \BDL.

For various reasons, different expansions of \BDL, often with 
non-classical connectives, have been studied.
The question arises whether the expansions whose connectives include one
or more non-classical connectives are interdefinable with an expansion
whose connectives include only classical connectives.
To investigate this question in a rigorous way, a precise definition of 
the interdefinability of propositional logics is needed.
It is natural to define the interdefinability of propositional logics in 
terms of the definability of connectives in a propositional logic.
However, it is not immediately clear how to define the definability of a 
connective in a propositional logic.   

Remarkably, almost all publications that refer to the definability of 
connectives in a propositional logic concern logics whose logical 
consequence relation is defined using a logical matrix and take the view 
that the definability of a connective is primarily a property of that 
logical matrix.
This view is appropriate to classical propositional logic provided its 
logical consequence relation is defined using the usual two-valued 
logical matrix.
However, it is not evident that it is appropriate to other propositional 
logics.
This means that, to investigate the above-mentioned question about the 
interdefinability of expansions of \BDL, it must first be determined  
what an appropriate definition of the definability of a connective is in 
the case of expansions of \BDL.

Fortunately, a general definition of the definability of a connective in 
a propositional logic, obtained by viewing it primarily as a property of 
the logical consequence relation of the logic, is given in~\cite{Smi62a} 
and elaborated on in~\cite{Woj88a}.
The definition that is adopted in this paper for \BDL\ and its 
expansions agrees with the definition given in~\cite{Smi62a}.
Because that definition seems to have been forgotten decades ago, it is 
discussed in this paper.
Due to the choice of the matrices used to define their logical 
consequence relations, a result from~\cite{Woj88a} also mentioned in 
that discussion provides a justification for the above view on the 
definability of a connective in the case of \BDL\ and its expansions.
This makes it relatively easy to establish the definability of a 
connective in \BDL\ and its expansions.

Expansions of \BDL\ that have been studied in earlier papers are usually 
expansions with one or more connectives that are not known from 
classical propositional logic.
Examples of such expansions are BD$\mathrm{\Delta}$~\cite{SO14a},
$\mathrm{E_{fde}}\!^{\Norm}$~\cite{CC20a}, F4CC~\cite{KZ20a}, and 
QLET$_\mathrm{F}$~\cite{ARCC22a}.  
Central to this paper is an expansion with connectives that are known 
from classical propositional logic, namely a falsity connective and an 
implication connective for which the standard deduction theorem holds.
This expansion will be referred to as $\BDL^{\IImpl,\False}$.
It has been treated in several earlier papers, 
including~\cite{AA96a,AA98a,AA17a,Pyn99a}, but without exception quite 
casually.

In most of this paper, the terms classical connective and non-classical 
connective are used rather loosely.
Roughly speaking, a connective of a non-classical logic is considered 
a classical connective if it has many properties in common with a 
connective known from classical logic, in particular properties with 
regard to logical consequence.
This paper also addresses the question to what extent the connectives of 
$\BDL^{\IImpl,\False}$ are classical connectives and the practical
consequences of the classical nature of the connectives of 
$\BDL^{\IImpl,\False}$.

The scope of this paper is limited to propositional logics equipped with 
a structural and non-trivial Scott consequence relation.
Precise definitions are required for various notions relevant to logics 
of the kind considered in this paper.
For most of the notions concerned, definitions can be found in the 
literature on logic.
However, those definitions are scattered over several publications and 
do not form a coherent whole (mainly due to differences in notation and 
terminology used).
Therefore, the relevant definitions are presented as a coherent whole in 
a preliminary section.
After its definition, a logic of the kind considered in this paper is  
simply called a logic. 

The structure of this paper is as follows.
First, preliminaries concerning propositional logics and logical 
matrices are presented (Section~\ref{PRELIMINARIES}).
Next, the general definitions of the definability of a connective in a 
propositional logic and the interdefinability of propositional logics 
used in subsequent sections are given (Sections~\ref{DEFINABILITY} and~\ref{INTERDEFINABILITY}).
After that, the language and logical consequence relation of \BDL\ are
introduced (Section~\ref{BELNAP-DUNN}). 
Then, the definability of connectives from expansions of \BDL\ and the
interdefinability of expansions of \BDL\ are investigated 
(Sections~\ref{DEFINABILITY-BD} and~\ref{INTERDEFINABILITY-BD}).
Thereafter, the question to what extent the connectives of 
$\BDL^{\IImpl,\False}$ are classical connectives and the practical
consequences of the classical nature of the connectives of 
$\BDL^{\IImpl,\False}$ are addressed
(Sections~\ref{CLASSICAL-LOGIC} and~\ref{MORE-CLASSICAL-LOGIC}).
Finally, some concluding remarks are made (Section~\ref{CONCLUSIONS}).

Old versions of~\cite{Mid23a} provide both a fairly comprehensive 
overview of the first-order version of $\BDL^{\IImpl,\False}$ and a 
study of the interdefinability of that logic with other expansions of 
the first-order version of Belnap-Dunn logic.
The current version of that paper only provides a revision of the 
overview of the first-order version of $\BDL^{\IImpl,\False}$.
This paper provides a major revision of the interdefinability study,  
considering only the propositional case.

\section{Logical Preliminaries}
\label{PRELIMINARIES}

This section briefly describes what a propositional logic is and how a 
propositional logic is defined using a logical matrix. 

\subsection{Propositional Logics}
\label{subsect-prelims-lang}

The language of a propositional logic is defined by way of an alphabet 
that consists of propositional variables and logical connectives.
\begin{sdef}
An \emph{alphabet} of a language of a propositional logic is a couple
$\LAlph = \tup{\PVar,\indfam{\Conn{n}}{n \in \Nat}}$, where:
\begin{itemize}
\item
$\PVar$ is a countably infinite set of \emph{propositional variables};
\item
$\indfam{\Conn{n}}{n \in \Nat}$ is an $\Nat$-indexed family of pairwise 
disjoint sets;
\item
for each $n \in \Nat$, $\Conn{n}$ is a finite set of 
\emph{logical connectives of arity $n$};
\item
$\Union \set{\Conn{n} \where n \in \Nat}$ is a non-empty finite set.
\end{itemize}
\end{sdef}

The language over an alphabet consists of formulas.
They are constructed according to the formation rules given below.

\begin{sdef}
Let $\LAlph = \tup{\PVar,\indfam{\Conn{n}}{n \in \Nat}}$ be an alphabet.
Then the set $\LForm^\LAlph$ of all \emph{formulas} over $\LAlph$ is 
inductively defined by the following formation rules:
\begin{itemize}
\item
if $p \in \PVar$, then $p \in \LForm^\LAlph$;
\item
if ${\Diamond} \in \Conn{0}$, then ${\Diamond} \in \LForm^\LAlph$;
\item
if ${\Diamond} \in \Conn{n+1}$ and
$A_1,\ldots,A_{n+1} \in \LForm^\LAlph$, then 
${\Diamond}(A_1,\ldots,A_{n+1}) \in \LForm^\LAlph$.
\end{itemize}
\pagebreak[2]
The set of all \emph{atomic formulas} over $\LAlph$, written 
$\LAForm^\LAlph$, is the set $\PVar$ of propositional variables.
\end{sdef}
The following will sometimes be used without mentioning (with or without 
decoration):
$p$ and $q$ as meta-variables ranging over all propositional variables 
from $\PVar^\LAlph$,
$A$ and $B$ as meta-variables ranging over all formulas from 
$\LForm^\LAlph$, and
$\Gamma$ and $\Delta$ as meta-variables ranging over all sets of 
formulas from $\LForm^\LAlph$.
We will write $\mathrm{var}(\Gamma)$, where $\Gamma$ is a set of 
formulas from $\LForm^\LAlph$, for the set of all propositional 
variables from $\PVar^\LAlph$ that occur in the formulas from $\Gamma$.

\begin{sdef}
Let $\LAlph = \tup{\PVar,\indfam{\Conn{n}}{n \in \Nat}}$ be an alphabet.
Then an \emph{$\LAlph$-substi\-tution} of formulas from $\LForm^\LAlph$ 
for variables from $\PVar$ is a function 
$\funct{\sigma}{\PVar}{\LForm^\LAlph}$.
An $\LAlph$-substitution $\sigma$ extends to the function 
$\funct{\sigma^*}{\LForm^\LAlph}{\LForm^\LAlph}$ that is 
recursively defined as follows:
\begin{ldispl}
\sigma^*(p) = \sigma(p),\; 
\sigma^*({\Box}) = {\Box},\;
\sigma^*({\Diamond}(A_1, \ldots, A_{n+1})) = 
 {\Diamond}(\sigma^*(A_1), \ldots,\sigma^*(A_{n+1})),
\end{ldispl}%
for ${\Box} \in \Conn{0}$ and ${\Diamond} \in \Conn{n+1}$.
\end{sdef}
We write $\sigma(A)$ for $\sigma^*(A)$ and
$\sigma(\Gamma)$ for $\set{\sigma^*(A) \where A \in \Gamma}$. 

We use the notational conventions to write $(\Diamond A)$ instead of 
$\Diamond(A)$ and $(A_1 \Diamond A_2)$ instead of $\Diamond(A_1,A_2)$ 
and to omit parenthesis where it does not lead to syntactic ambiguities 
if the previous convention is used.

\begin{sdef}
Let $\LAlph = \tup{\PVar,\indfam{\Conn{n}}{n \in \Nat}}$ be an alphabet.
Then a \emph{logical consequence relation for $\LForm^\LAlph$} is a 
binary relation $\LCon$ on $\pset(\LForm^\LAlph)$ that satisfies the 
following conditions:%
\footnote
{As usual, we write $\Gamma,\Gamma'$ for $\Gamma \union \Gamma'$ and $A$
 for $\set{A}$.}
\begin{itemize}
\item
if $\Gamma \inter \Delta \neq \emptyset$ then $\Gamma \LCon \Delta$;
\item
if $\Gamma \LCon \Delta$, $\Gamma \subseteq \Gamma'$, and 
$\Delta \subseteq \Delta'$ then $\Gamma' \LCon \Delta'$;
\item
if $\Gamma \LCon \Delta,A$ and $A,\Gamma' \LCon \Delta'$ then 
$\Gamma,\Gamma' \LCon \Delta,\Delta'$.
\end{itemize}
\end{sdef}

\pagebreak[2]
\begin{sdef}
A \emph{(propositional) logic} is a couple $(\LAlph,\LCon)$, where:
\begin{itemize}
\item
$\LAlph$ is an alphabet;
\item
$\LCon$ is a logical consequence relation for $\LForm^\LAlph$ that
satisfies the following conditions:

\begin{itemize}
\item
if $\Gamma \LCon \Delta$ and $\sigma$ is an $\LAlph$-substitution then 
$\sigma(\Gamma) \LCon \sigma(\Delta)$;
\item
there exist non-empty $\Gamma$ and $\Delta$ such that not 
$\Gamma \LCon \Delta$.
\end{itemize}
\end{itemize}
A \emph{finitary logic} is a logic $(\LAlph,\LCon)$ where $\LCon$ is 
such that:
\begin{itemize}
\item[]
if $\Gamma \LCon \Delta$ then there exists finite 
$\Gamma' \subseteq \Gamma$ and $\Delta' \subseteq \Delta$ such that 
$\Gamma' \LCon \Delta'$.
\end{itemize}
A \emph{uniform logic} is a logic $(\LAlph,\LCon)$  where $\LCon$ is 
such that:
\begin{itemize}
\item[]
if $\Gamma,\Gamma' \LCon \Delta,\Delta'$ and
$\mathrm{var}(\Gamma \union \Delta) \inter
 \mathrm{var}(\Gamma' \union \Delta') = \emptyset$
then $\Gamma \LCon \Delta$ or $\Gamma' \LCon \Delta'$.
\end{itemize}
\end{sdef}

\subsection{Logical Matrices}

The interpretation of the logical connectives of a logic can often be 
given using a logical matrix.
\begin{sdef}
Let $\LAlph = \tup{\PVar,\indfam{\Conn{n}}{n \in \Nat}}$ be an alphabet.
Then a \emph{(logical) matrix for $\LAlph$} is a triple 
$(\TValue, \DValue, \TFunct)$, where:
\pagebreak[2]
\begin{itemize}
\item
$\TValue$ is a non-empty set of \emph{truth values};
\item
$\DValue \subset \TValue$ is a non-empty set of 
\emph{designated truth values};
\item
$\TFunct$ is a function from $\Union \set{\Conn{n} \where n \in \Nat}$ 
to $\Union \set{\funct{f}{\TValue^n}{\TValue} \where n \in \Nat}$ such 
that, for each $n \in \Nat$, for each $\Diamond \in \Conn{n}$, 
$\funct{\TFunct(\Diamond)}{\TValue^n}{\TValue}$.
\end{itemize}
A \emph{finite matrix} is a matrix $(\TValue,\DValue,\TFunct)$ where 
$\TValue$ is a finite set. 
A $n$-\emph{valued matrix} is a finite matrix 
$(\TValue,\DValue,\TFunct)$ where the cardinality of $\TValue$ is $n$. 
\\[1ex]
The set of \emph{non-designated truth values} of a matrix 
$(\TValue,\DValue,\TFunct)$, written $\NDValue$, is 
$\TValue \diff \DValue$. 
\end{sdef}
For an alphabet $\LAlph = \tup{\PVar,\indfam{\Conn{n}}{n \in \Nat}}$ and 
a matrix $\Matrix = \langle \TValue, \DValue, \TFunct \rangle$ for 
$\LAlph$, a valuation of the formulas from $\LForm^\LAlph$ in $\Matrix$ 
is given by a function that maps each formula from $\LForm^\LAlph$ to an 
element of $\TValue$ according to the principle of compositionality.
\begin{sdef}
\label{def-valuation}
\sloppy
Let $\LAlph = \tup{\PVar,\indfam{\Conn{n}}{n \in \Nat}}$ be an alphabet,
and let $\Matrix = (\TValue, \DValue, \TFunct)$ be the matrix for 
$\LAlph$.
Then a \emph{truth-functional valuation in $\Matrix$} is a function
$\funct{\nu}{\LForm^\LAlph}{\TValue}$ that satisfies the following
conditions:
\begin{itemize}
\item
if ${\Diamond} \in \Conn{0}^\LAlph$ then 
$\LVal{{\Diamond}} = \TFunct({\Diamond})$;
\item
if ${\Diamond} \in \Conn{n+1}^\LAlph$ and 
$A_1,\ldots,A_{n+1} \in \LForm^\LAlph$ then
$\LVal{{\Diamond}(A_1,\ldots,A_{n+1})} =
 \TFunct({\Diamond})(\LVal{A_1},\ldots,\LVal{A_{n+1}})$.%
\footnote
{Logics induced by a matrix using as valuations functions from
 $\LForm^\LAlph$ to $\TValue$ that are not truth-functional valuations
 are not considered in this paper.}
\end{itemize}
\end{sdef}
In the sequel, a truth-functional valuation is simply called a 
valuation.

A matrix for some alphabet induces a logical consequence relation.
\begin{sdef}
\label{def-LCon}
Let $\LAlph$ be an alphabet, and 
let $\Matrix = (\TValue, \DValue, \TFunct)$ be the matrix for~$\LAlph$.
Then the \emph{logical consequence relation induced by $\Matrix$} is 
the logical consequence relation $\LCon_\Matrix$ for $\LForm^\LAlph$
that is defined as follows:
\begin{ldispl}
\Gamma \mathrel{\LCon_\Matrix} \Delta \;\mathrm{iff}\;
\mathrm{for\; all\; valuations}\; \nu \;\mathrm{in}\; \Matrix, 
\\ 
\phantom{\Gamma \mathrel{\LCon_\Matrix} \Delta \;\mathrm{iff}\;} 
\quad\;\,
\mathrm{if}\; 
\LVal{A} \in \DValue  \;\mathrm{for\; all}\;  A  \in \Gamma
\;\mathrm{then}\; 
\LVal{A'} \in \DValue \;\mathrm{for\; some}\; A' \in \Delta\;.
\end{ldispl}%
\end{sdef}
The following theorems are well-known results about matrices
(see e.g.~\cite{Woj88a}, Theorems~3.2.5 and~3.2.7).
\begin{theorem}
Let $\LAlph$ be an alphabet, and 
let $\Matrix$ be a matrix for $\LAlph$.
Then:
\begin{itemize}
\item
$(\LAlph,\LCon_\Matrix)$ is a uniform logic;
\item
if $\Matrix$ is a finite matrix then $(\LAlph,\LCon_\Matrix)$ is a 
finitary and uniform logic.
\end{itemize}
\end{theorem}

\begin{theorem}
Let $(\LAlph,\LCon)$ be a finitary and uniform logic.
Then there exists a finite matrix $\Matrix$ for $\LAlph$ such that
$\LCon \;=\; \LCon_\Matrix$.
\end{theorem}

Some definitions and results to come refer to matrix functions and 
simple matrices.
\begin{sdef}
\label{def-matrix-function}
Let $\Matrix = (\TValue,\DValue,\TFunct)$ be a matrix.
Then a \emph{matrix function of $\Matrix$} is a function 
$\funct{f}{\TValue^n}{\TValue}$ ($n \in \Nat$) obtainable by 
composition from the functions in the image of $\TFunct$ and the 
projection functions on finite cartesian powers of $\TValue$.
\end{sdef}
\begin{sdef}
\label{def-simple-matrix}
Let $\Matrix = (\TValue,\DValue,\TFunct)$ be a matrix.
Then $\Matrix$ is a \emph{simple matrix} iff, for all $n \in \Nat$, 
for all $a_1,\ldots,a_n,b_1,\ldots,b_n \in \TValue$, 
if, for all $n$-ary matrix functions $f$ of $\Matrix$,
$f(a_1,\ldots,a_n) \in \DValue$ iff 
$f(b_1,\ldots,b_n) \in \nolinebreak \DValue$,
then $a_1 = b_1$, \ldots, $a_n = b_n$.
\end{sdef}

The following is a corollary of Definitions~\ref{def-valuation} 
and~\ref{def-matrix-function}.
\begin{corollary}
\label{corollary-matrix-function}
Let $\LAlph$ be an alphabet,
let $\Matrix = (\TValue,\DValue,\TFunct)$ be a matrix for $\LAlph$, and 
let $\funct{f}{\TValue^n}{\TValue}$ ($n \in \Nat$).
Then $f$ is a matrix function of $\Matrix$ iff
there exists a formula $A \in \LForm^\LAlph$ and propositional variables
$p_1,\ldots,p_n \in \PVar$ such that
$p_1,\ldots,p_n$ are the distinct propositional variables occurring in 
$A$ and
for all valuations $\nu$ in $\Matrix$,
$f(\nu(p_1),\ldots,\nu(p_n)) = \nu(A)$.
\end{corollary}

A matrix that is not a simple matrix has distinct truth values that are 
indistinguishable and therefore identifiable with each other.
For this reason, such matrices must be excluded in a useful definition 
of an $n$-valued logic.
\begin{sdef}
An \emph{$n$-valued logic} ($n \geq 2$) is a logic $(\LAlph,\LCon)$ 
where $\LCon \;=\; \LCon_\Matrix$ for some $n$-valued simple matrix 
$\Matrix = (\TValue,\DValue,\TFunct)$ for $\LAlph$.
\end{sdef}

The proof of some results to come refer to truth-functional completeness.
\begin{sdef}
\label{def-truth-func-compl}
Let $\Logic = (\LAlph,\LCon)$ be an $n$-valued logic.
Then \emph{$\Logic$ is truth-func\-tionally complete} iff, for some 
$n$-valued simple matrix $\Matrix = (\TValue, \DValue, \TFunct)$ such 
that $\LCon \;=\; \LCon_\Matrix$, for each $m \in \Nat$, for each 
$\funct{f}{\TValue^m}{\TValue}$, $f$ is a matrix function of $\Matrix$.
\end{sdef}

A matrix for some alphabet also induce a logical equivalence relation.
\begin{sdef}
\label{def-LEqv}
Let $\LAlph$ be an alphabet, and 
let $\Matrix = (\TValue, \DValue, \TFunct)$ be the matrix for~$\LAlph$.
Then the \emph{logical equivalence relation induced by $\Matrix$} is
the equivalence relation $\LEqv_\Matrix$ on $\LForm^\LAlph$ that is 
defined as follows:
\begin{ldispl}
A_1 \mathrel{\LEqv_\Matrix} A_2 \;\mathrm{iff}\;
\mathrm{for\; all\; valuations}\; \nu \;\mathrm{in}\; \Matrix,
\LVal{A_1} = \LVal{A_2}\;.
\end{ldispl}
\end{sdef}
It holds that $A_1 \mathrel{\LEqv_\Matrix} A_2$ only if 
$A_1 \mathrel{\LCon_\Matrix} A_2$ and $A_2 \mathrel{\LCon_\Matrix} A_1$.
In general, it does not hold that $A_1 \mathrel{\LEqv_\Matrix} A_2$ if 
$A_1 \mathrel{\LCon_\Matrix} A_2$ and $A_2 \mathrel{\LCon_\Matrix} A_1$.
However, it does hold if $\Matrix$ is the simple matrix that induces the 
logical consequence relation of a version of classical propositional 
logic.

\section{Synonymity and Definability of Connectives}
\label{DEFINABILITY}

What does it mean that a connective is definable in a logic?
If the definability of connectives is treated in publications on 
classical logic, it is usually defined as follows: 
an $n$-ary connective $\Diamond$ is definable iff 
${\Diamond}(p_1,\ldots,p_n) \LEqv A$ for some formula $A$ in which 
$\Diamond$ does not occur.
A justification of this definition is almost always lacking.
An obvious justification is that, if $\Diamond$ is definable, it need 
not be regarded as basic because there is a formula $A$ such that, in 
each formula in which one or more subformulas of the form 
${\Diamond}(A_1,\ldots,A_n)$ occur, these occurrences may always be 
replaced by appropriate substitution instances of $A$.
In the case of classical logic, a connective has this replaceability
property iff it is definable according to the definition in terms of
the logical equivalence relation given above.
However, it is not clear whether this is the case in general.

If a formula may always be replaced by another formula, then those
formulas are called synonymous.
A definition of synonymity can be given for an arbitrary logic solely in 
terms of its constituent parts, i.e.\ its language and its logical 
consequence relation.
\begin{sdef}
\label{def-LSyn}
Let $\Logic = (\LAlph,\LCon)$ be a logic. 
Then the \emph{synonymity relation of $\Logic$} is the equivalence 
relation $\LSyn_\Logic$ on $\LForm^\LAlph$ that is defined as follows:
\begin{itemize}
\item[]
$A_1 \LSyn_\Logic A_2$ iff,
for all formulas $B_1$ and $B_2$ from $\LForm^\LAlph$ such that 
$B_2$ is $B_1$ with some or all occurrences of $A_1$ replaced by $A_2$, 
$B_1 \LCon_\Logic B_2$ and $B_2 \LCon_\Logic B_1$.
\end{itemize}
\end{sdef}
It is easy to see that $\LSyn_\Logic$, where $\Logic = (\LAlph,\LCon)$, 
is an equivalence relation on $\LForm^\LAlph$.

The logical equivalence relation induced by a matrix is included in the
synonymity relation of the logic induced by that matrix.
The question arises whether the reverse is also the case.
In general, this question cannot be answered in the affirmative.
However, for all logics whose logical consequence relation is induced by 
a simple matrix, the question can be answered in the affirmative.

\begin{theorem}
\label{theorem-lsyn-leqv-gen}
Let $\LAlph$ be an alphabet,
let $\Matrix = (\TValue,\DValue,\TFunct)$ be a simple matrix for 
$\LAlph$, and
let $\Logic = (\LAlph,\LCon_\Matrix)$.
Then, for all $A_1,A_2 \in \LForm^\LAlph$, $A_1 \LSyn_\Logic A_2$ iff
$A_1 \LEqv_\Matrix A_2$.
\end{theorem}
\begin{proof}
This follows immediately from the definition of simple matrices 
(Definition~\ref{def-simple-matrix}) and Lemma~16.11 in~\cite{SS78a}.
\qed
\end{proof}
Put in plain language, a simple matrix is a matrix in which different 
truth values can always be distinguished.
In~\cite{Smi62a}, a matrix with three truth values is presented that 
induces the logical consequence relation of a version of classical 
propositional logic.
In that matrix, there are two truth values that cannot be distinguished
and the induced logical equivalence relation does not include the
synonymity relation of the version of classical propositional logic in
question.
The matrix concerned is also an example of the phenomenon that the 
logical consequence relation of an $n$-valued logic (as defined in this 
paper) is also induced by non-simple matrices with more than $n$ truth 
values.

A general definition of the definability of a connective in a logic that 
has the justification mentioned in the first paragraph of this section 
can be easily given using the synonymity relation of the logic.
\begin{sdef}
\label{def-definability}
Let $\LAlph = \tup{\PVar,\indfam{\Conn{n}}{n \in \Nat}}$ be an alphabet,
let $\Logic = (\LAlph,\LCon)$ be a logic, 
let $\Diamond \in \Conn{n}$ ($n \in \Nat$), and
let $C \subseteq
     \Union \set{\Conn{n} \where n \in \Nat} \diff \set{\Diamond}$.
Then \emph{$\Diamond$ is definable in $\Logic$ in terms of $C$} 
iff there exist $p_1,\ldots,p_n \in \PVar$ and an $A \in \LForm^\LAlph$ 
in which only logical connectives from $C$ occur such that 
${\Diamond}(p_1,\ldots,p_n) \LSyn_\Logic A$.
%
%
We say that \emph{$\Diamond$ is definable in $\Logic$} if there exists a
$C \subseteq
 \Union \set{\Conn{n} \where n \in \Nat} \diff \set{\Diamond}$ 
such that $\Diamond$ is definable in $\Logic$ in terms of $C$.
\end{sdef}
Synonymity of formulas also underlies the proposal made in~\cite{Jur87a} 
to use two-way inference rules as definitions of connectives. 

When publications on multi-valued logics state that a certain connective 
is definable, it is usually not made precise what is meant by the 
definability of a connective.
A recent exception is~\cite{SO14a}.
In that paper, it is made precise what is meant by the definability of a 
connective in an expansion of Belnap-Dunn logic.
The relation that must hold between ${\Diamond}(p_1,\ldots,p_n)$ and the 
defining formula~$A$ is defined in terms of the logical consequence 
relation of the expansion.
No justification is given for that definition.
However, the relation concerned coincides with the logical equivalence
relation induced by the simple matrix that induces the logical 
consequence relation of the expansion, and therefore by 
Theorem~\ref{theorem-lsyn-leqv-gen} also with the synonymity relation of 
the expansion (see also the last paragraph of 
Section~\ref{BELNAP-DUNN}).

The definition of the synonymity relation of a logic given above 
corresponds to the definitions given in~\cite{Smi62a,SS78a}.
The definition of the definability of a connective in a logic given 
above corresponds to the definitions given in~\cite{Smi62a,Woj88a}.

\section{Expansions and Interdefinability of Logics}
\label{INTERDEFINABILITY}

Using the definition of the definability of a connective in a logic 
given in Section~\ref{DEFINABILITY}, the interdefinability of two 
$n$-valued logics ($n \geq 2$) can be easily defined by involving a 
third logic that expands the two logics.
\begin{sdef}
Let $\LAlph = \tup{\PVar,\indfam{\Conn{n}}{n \in \Nat}}$ and
$\LAlph' = \tup{\PVar',\indfam{\Conn{n}'}{n \in \Nat}}$ be alphabets 
such that $\PVar = \PVar'$ and $\Conn{n} \subseteq \Conn{n}'$ for each 
$n \in \Nat$, and
let $\Matrix = (\TValue, \DValue, \TFunct)$ and
$\Matrix' = (\TValue', \DValue', \TFunct')$ 
be matrices for $\LAlph$ and $\LAlph'$, respectively. 
Then \emph{$\Matrix'$ is an expansion of $\Matrix$} if 
$\TValue = \TValue'$, $\DValue = \DValue'$, and 
$\TFunct(\Diamond) = \TFunct'(\Diamond)$ for each 
$\Diamond \in \Union \set{\Conn{n} \where n \in \Nat}$.
\end{sdef}

\begin{sdef}
\label{def-expansion-logic}
Let $\Logic = (\LAlph,\LCon)$ and $\Logic' = (\LAlph',\LCon')$ be 
$n$-valued logics ($n \geq 2$).
Then \emph{$\Logic'$ is an expansion of $\Logic$} if there exist a 
matrix $\Matrix$ for $\LAlph$ and a matrix $\Matrix'$ for $\LAlph'$ such 
that $\LCon \;=\; \LCon_\Matrix$, $\LCon' \;=\; \LCon_{\Matrix'}$, and
$\Matrix'$ is an expansion of $\Matrix$.
\end{sdef}
As a corollary of Definition~\ref{def-expansion-logic}, we have that, 
for all $n$-valued logics $\Logic = (\LAlph,\LCon)$ and 
$\Logic' = (\LAlph',\LCon')$, $\Logic'$ is an expansion of $\Logic$ iff 
$\Logic'$ is a conservative extension of $\Logic$, i.e., for all 
$\Gamma,\Delta \subseteq \LForm^\LAlph$, $\Gamma \LCon \Delta$ 
iff $\Gamma \LCon' \Delta$.

In the coming sections, we use a special notation for referring to 
expansions of logics.

Let $\Logic = (\LAlph,\LCon_\Matrix)$ be a uniform logic, where
$\LAlph = (\PVar,\indfam{\Conn{n}}{n \in \Nat})$ and
$\Matrix = (\TValue,\DValue,\TFunct)$.
Moreover, let the following be given: 
\begin{itemize}
\item
connectives $\Diamond_1,\ldots,\Diamond_m$ such that 
$\set{\Diamond_1,\ldots,\Diamond_m} \inter
 \Union \set{\Conn{n}\where n \in \Nat} = \emptyset$;
\item
an arity $n_i$ of $\Diamond_i$ for each $i \in \set{1,\ldots,m}$;
\item
an intended interpretation 
$\funct{\widehat{\Diamond}_i}{\TValue^{n_i}}{\TValue}$ of
$\Diamond_i$ for each $i \in \set{1,\ldots,m}$.
\end{itemize}
Then we write $\Logic^{\Diamond_1,\ldots,\Diamond_m}$ for the expansion 
$\Logic' = (\LAlph',\LCon_{\Matrix'})$ of $\Logic$, where:
\begin{itemize}
\item 
$\LAlph' = (\PVar,\indfam{\Conn{n}'}{n \in \Nat})$, 
where
\begin{itemize}
\item[]
$\Conn{n}' = \Conn{n} \union \set{\Diamond_i \where n_i = n}$
for each $n \in \Nat$;
\end{itemize}
\item
$\Matrix' = (\TValue,\DValue,\TFunct')$,
where $\TFunct'$ is defined as follows:
\begin{itemize}
\item[]
$\TFunct'(\Diamond) = \TFunct(\Diamond)$ 
if $\Diamond \in \Union \set{\Conn{n} \where n \in \Nat}$,
\item[]
$\TFunct'(\Diamond) = \widehat{\Diamond}$ \hspace*{1.25em}
if $\Diamond \in \set{\Diamond_1,\ldots,\Diamond_m}$.
\end{itemize}
\end{itemize}

The following corollary of Corollary~\ref{corollary-matrix-function}, 
Theorem~\ref{theorem-lsyn-leqv-gen}, and 
Definitions~\ref{def-truth-func-compl}, \ref{def-definability}, 
and~\ref{def-expansion-logic} relates truth-functional completeness to 
definability of connectives.
\begin{corollary}
\label{corollary-truth-func-compl}
Let $\LAlph = (\PVar,\indfam{\Conn{n}}{n \in \Nat})$ be an alphabet,
let $\Matrix = (\TValue,\DValue,\TFunct)$ be an $n$-valued simple matrix 
for $\LAlph$ ($n \in \Nat$), and
let $\Logic = (\LAlph,\LCon_\Matrix)$.
Moreover, \linebreak[2] let, for all 
$f \in \Union \set{\funct{f}{\TValue^n}{\TValue} \where n \in \Nat}$,
$\Diamond_f$ be a connective for which it is given that the intended 
interpretation is $f$.
Then $\Logic$ is truth-functionally complete iff for all 
$f \in \Union \set{\funct{f}{\TValue^n}{\TValue} \where n \in \Nat}$ 
with $\Diamond_f \notin \Union \set{\Conn{n}\where n \in \Nat}$, 
$\Diamond_f$ is definable in $\Logic^{\Diamond_f}$.
\end{corollary}

Two ($n$-valued) logics are interdefinable if they have a common 
expansion in which the connectives of each of them are definable in 
terms of the connectives of the other one.
\begin{sdef}
Let $\LAlph = (\PVar,\indfam{\Conn{n}}{n \in \Nat})$ and
$\LAlph' = (\PVar',\indfam{\Conn{n}'}{n \in \Nat})$ be alphabets such
that $\PVar = \PVar'$, and
let $\Logic = (\LAlph,\LCon)$ and $\Logic' = (\LAlph',\LCon')$ be 
$n$-valued logics ($n \geq 2$). 
Then \emph{$\Logic$ is definable in $\Logic'$} iff
there exists a logic $\Logic''$ such that $\Logic''$ is an expansion of
both $\Logic$ and $\Logic'$ and each 
$\Diamond \in \Union \set{\Conn{n} \where n \in \Nat}$ is definable in 
$\Logic''$ in terms of $\Union \set{\Conn{n}' \where n \in \Nat}$.
\emph{$\Logic$ is interdefinable with $\Logic'$}, written 
$\Logic \simeq \Logic'$, iff \linebreak[2] $\Logic$ is definable in 
$\Logic'$ and $\Logic'$ is definable in $\Logic$.
\end{sdef}
It is easy to see that $\simeq$ is an equivalence relation on 
$n$-valued logics.

\section{Belnap-Dunn Logic}
\label{BELNAP-DUNN}

Belnap-Dunn logic (\BDL) and its expansions are $4$-valued logics that 
have been studied relatively extensively.
Below, the language of \BDL\ and the logical consequence relation of 
\BDL\ are concisely introduced.

\pagebreak[2]
The language of \BDL\ is defined by way of its alphabet.

\begin{sdef}
\label{def-LAlph-BD}
The alphabet $\LAlph^\BDL$ of the language of an instance of \BDL\ is 
a couple $(\PVar^\BDL,\indfam{\Conn{n}^\BDL}{n \in \Nat})$, where:
\begin{itemize}
\item
$\PVar^\BDL$ is a countably infinite set of propositional variables;
\item
$\Conn{1}^\BDL = \set{\Not}$;
\item
$\Conn{2}^\BDL = \set{\CAnd,\COr}$; 
\item
$\Conn{n+3}^\BDL = \emptyset$ for each $n \in \Nat$.
\end{itemize}
Each choice of $\PVar^\BDL$ gives rise to a different instance of 
\BDL. 
In this paper, a fixed but arbitrary choice of $\PVar^\BDL$ is 
assumed for \BDL\ and all expansions of BD.
\end{sdef}
We write $\LForm^\BDL$ and $\LAForm^\BDL$ instead of 
$\LForm^{\LAlph^\BDL}$ and $\LAForm^{\LAlph^\BDL}$, respectively.
The superscript $\BDL$ will be omitted from $\PVar^\BDL$, 
$\Conn{n}^\BDL$, $\LAlph^\BDL$, $\LForm^\BDL$, and $\LAForm^\BDL$ if no 
confusion can arise.

The logical consequence relation of \BDL\ is induced by a matrix. 
In the definition of this matrix, $\VTrue$ (\emph{true only}), 
$\VFalse$ (\emph{false only}), $\VBoth$ (\emph{both true and false}), 
and $\VNeither$ (\emph{neither true nor false}) are taken as truth 
values.
Moreover, use is made of the partial order $\leq$ on the set 
$\set{\VTrue,\VFalse,\VBoth,\VNeither}$ in which $\VFalse$ is the least 
element, $\VTrue$ is the greatest element, and $\VBoth$ and $\VNeither$ 
are incomparable.
We write $\inf V$ and $\sup V$, 
where $V \subseteq \set{\VTrue,\VFalse,\VBoth,\VNeither}$, for the
greatest lower bound and least upper bound, respectively, of $V$ with 
respect to $\leq$.
\begin{sdef}
\label{def-Matrix-BD}
The matrix $\Matrix^\BDL$ for $\LAlph^\BDL$ is the triple 
$(\TValue^\BDL,\DValue^\BDL,\TFunct^\BDL)$, where:
\begin{itemize}
\item
$\TValue^\BDL = \set{\VTrue,\VFalse,\VBoth,\VNeither}$;
\item
$\DValue^\BDL = \set{\VTrue,\VBoth}$;
\item
$\TFunct^\BDL$ is defined as follows:
\[
\begin{array}{rcl}
\TFunct^\BDL(\Not)(a) & = &
 \left \{
 \begin{array}{l@{\;\;}l}
 \VTrue  & \mathrm{if}\; a = \VFalse \\
 \VFalse & \mathrm{if}\; a = \VTrue \\
 a       & \mathrm{otherwise}\;,
 \end{array}
 \right.
\vspace*{.5ex} \\
\TFunct^\BDL(\CAnd)(a_1,a_2) & = &
 \begin{array}[t]{l}
 \inf \set{a_1,a_2}\;,
 \end{array}
\vspace*{.5ex} \\
\TFunct^\BDL(\COr)(a_1,a_2) & = &
 \begin{array}[t]{l}
 \sup \set{a_1,a_2}\;,
 \end{array}
\end{array}
\]
where 
$a$, $a_1$, and $a_2$ range over all truth values from $\TValue^\BDL$.
\end{itemize}
\end{sdef}
We write $\LCon_\BDL$ and $\LEqv_\BDL$ instead of 
$\LCon_{\Matrix^\BDL}$ and $\LEqv_{\Matrix^\BDL}$, respectively.
The superscript or subscript $\BDL$ will be omitted from 
$\TValue^\BDL$, $\DValue^\BDL$, $\TFunct^\BDL$, 
$\Matrix^\BDL$, $\LCon_\BDL$, $\LEqv_\BDL$, and $\LSyn_\BDL$ 
if no confusion can arise.

Moreover, we write \CL\ for the version of classical propositional logic 
with the same alphabet as \BDL\ and \CLp\ for the positive fragment of 
\CL.

The submatrix of $\Matrix^\BDL$ induced by restriction of the set of 
truth values to $\set{\VTrue,\VFalse,\VBoth}$ is the matrix inducing the
logical consequence relation $\LCon_\LP$ of Priest's logic of 
paradox (\LP)~\cite{Pri79a}.
The submatrix of $\Matrix^\BDL$ induced by restriction of the set of 
truth values to $\set{\VTrue,\VFalse,\VNeither}$ is the matrix inducing 
the logical consequence relation $\LCon_\Kiii$ of Kleene's strong 
three-valued logic (\Kiii)~\cite{Kle52a}.
The submatrix of $\Matrix^\BDL$ induced by restriction of the set of 
truth values to $\set{\VTrue,\VFalse}$ is the matrix inducing the
logical consequence relation $\LCon_\CL$.
From this, it follows easily that
${\LCon_\BDL} \subset {\LCon_\LP}\subset {\LCon_\CL}$ and
${\LCon_\BDL} \subset {\LCon_\Kiii}\subset {\LCon_\CL}$.
\LP, \Kiii, and \CL\ are extensions of \BDL: they have the 
same language as \BDL\ and their logical consequence relations 
include the logical consequence relation of \BDL.

In the case of \BDL, the logical equivalence relation can be defined in 
terms of the logical consequence relation. 
\begin{theorem}
\label{theorem-leqv-lcon-bd}
\!For all $A_1,A_2 \in \LForm^\BDL$, $ A_1 \LEqv_\BDL A_2$ iff
$A_1 \LCon_\BDL A_2$, $A_2 \LCon_\BDL \nolinebreak A_1$, 
$\Not A_1 \LCon_\BDL \Not A_2$, and $\Not A_2 \LCon_\BDL \Not A_1$.
\end{theorem}
\begin{proof}
This follows easily from Definitions~\ref{def-LCon}, \ref{def-LEqv}, 
\ref{def-LAlph-BD}, and~\ref{def-Matrix-BD}.
\qed
\end{proof}

We have that logical equivalence implies synonymity.
In the case of \BDL, we have in addition that synonymity implies 
logical equivalence.
\begin{theorem}
\label{theorem-lsyn-leqv-bd}
For all $A_1,A_2 \in \LForm^\BDL$, $A_1 \LSyn_\BDL A_2$ iff
$A_1 \LEqv_\BDL A_2$.
\end{theorem}
\begin{proof}
Let $a_1,a_2 \in \TValue$.
Assume that $a_1 \neq a_2$.
In the case where $a_1,a_2 \in \DValue$ or $a_1,a_2 \in \NDValue$, 
we have that
$\TFunct(\Not)(a_1) \in \DValue$ iff $\TFunct(\Not)(a_2) \in \NDValue$.
In all other cases, we trivially have that $a_1 \in \DValue$ iff 
$a_2 \in \NDValue$.
Hence, if $a_1 \neq a_2$ then there exists a matrix function $f$ such 
that $f(a_1) \in \DValue$ iff $f(a_2) \in \NDValue$.
This means that $\Matrix$ is a simple matrix.
From this and Theorem~\ref{theorem-lsyn-leqv-gen} it follows that, for 
all $A_1,A_2 \in \LForm$, $A_1 \LSyn A_2$ iff $A_1 \LEqv A_2$.
\qed
\end{proof}
As a corollary of Theorems~\ref{theorem-leqv-lcon-bd} 
and~\ref{theorem-lsyn-leqv-bd}, we have that, 
for all $A_1,A_2 \in \LForm^\BDL$, $ A_1 \LSyn_\BDL A_2$ iff
$A_1 \LCon_\BDL A_2$, $A_2 \LCon_\BDL \nolinebreak A_1$, 
$\Not A_1 \LCon_\BDL \Not A_2$, and $\Not A_2 \LCon_\BDL \Not A_1$.

It follows immediately from the proofs of 
Theorems~\ref{theorem-leqv-lcon-bd} and~\ref{theorem-lsyn-leqv-bd} that
these theorems go through for expansions of \BDL.
Consequently, the above-mentioned corollary goes through for expansions 
of \BDL\ too.
This justifies the definition of the definability of a 
connective in an expansion of Belnap-Dunn logic given in~\cite{SO14a}
(and mentioned in Section~\ref{DEFINABILITY}).

\section{Definability of Connectives from Expansions of \BDL}
\label{DEFINABILITY-BD}

Well-known classical connectives with which \BDL\ can be expanded are
among other things the nullary \emph{falsity connective} $\False$ and 
the binary \emph{implication connective} $\IImpl$.
\begin{sdef}
\label{def-class-conn}
The intended interpretation of the classical connectives $\False$ and 
$\IImpl$ are the functions $\funct{\widehat{\False}}{\,}{\TValue}$ and 
$\funct{\widehat{\IImpl}}{\TValue \x \TValue}{\TValue}$, respectively, 
where $\TValue = \set{\VTrue,\VFalse,\VBoth,\VNeither}$, defined as 
follows: 
\[
\renewcommand{\arraystretch}{1.25}
\begin{array}[t]{rcl}
\widehat{\False} & = &
 \begin{array}[t]{l}
 \VFalse \;,
 \end{array}
\\[.5ex]
\widehat{\hspace*{.1em}\IImpl}(a_1,a_2) & = &
 \left \{
 \begin{array}{l@{\;\;}l}
 \VTrue  & \mathrm{if}\; a_1 \notin \set{\VTrue,\VBoth} \\
 a_2 & \mathrm{otherwise}\;.
 \end{array}
 \right.
\end{array}
\]
\end{sdef}

Several interesting non-classical connectives have been added to \BDL\ 
in studies of expansions of \BDL\ (or a first-order version of \BDL).
Among them are
the unary \emph{is-designated connective} $\Des$, 
the unary \emph{classicality connective} $\Norm$, 
the unary \emph{consistency connective} $\Cons$,  
the unary \emph{determinedness connective} $\Det$, and
the unary \emph{conflation connective} $\Confl$.
The connective $\Des$ is for example found in the logic 
BD$\mathrm{\Delta}$ studied in~\cite{SO14a} and 
the connective $\Norm$ is for example found in the logic 
QLET$_\mathrm{F}$ studied in~\cite{ARCC22a}.%
\footnote
{In~\cite{ARCC22a}, the symbol $\Cons$ is used instead of $\Norm$.
 The symbol $\Norm$ is taken from~\cite{CC20a}.}
The consistency connective $\Cons$ and the determinedness connective 
$\Det$ are for example discussed in~\cite{CC20a}.
In the setting of Priest's logic of paradox, the consistency connective 
is found in the logic $\LP^{\Cons}$~\cite{Pic18a}. 
\begin{sdef}
\label{def-nonclass-conn}
The intended interpretation of the non-classical connectives $\Des$, 
$\Norm$, $\Cons$, $\Det$, and $\Confl$ are the functions 
$\funct{\widehat{\Des}}{\TValue}{\TValue}$,\, 
$\funct{\widehat{\Norm}}{\TValue}{\TValue}$,\, 
$\funct{\widehat{\Cons}}{\TValue}{\TValue}$, 
$\funct{\widehat{\Det}}{\TValue}{\TValue}$, and
$\funct{\widehat{\Confl}}{\TValue}{\TValue}$, respectively, 
where  $\TValue = \set{\VTrue,\VFalse,\VBoth,\VNeither}$, defined as 
follows: 
\[
\begin{array}[t]{r@{\;\;}c@{\;\;}l}
\widehat{\Des}(a) & = &
 \left \{
 \begin{array}{l@{\;\;}l}
 \VTrue  & \mathrm{if}\; a \in \set{\VTrue,\VBoth} \\
 \VFalse & \mathrm{otherwise},
 \end{array}
 \right.
\vspace*{.5ex} \\
\widehat{\Norm}(a) & = &
 \left \{
 \begin{array}{l@{\;\;}l}
 \VTrue  & \mathrm{if}\; a\in \set{\VTrue,\VFalse} \\
 \VFalse & \mathrm{otherwise},
 \end{array}
 \right.
\vspace*{.5ex} \\
\widehat{\Cons}(a) & = &
 \left \{
 \begin{array}{l@{\;\;}l}
 \VTrue  & \mathrm{if}\; a \in \set{\VTrue,\VFalse,\VNeither} \\
 \VFalse & \mathrm{otherwise},
 \end{array}
 \right.
\vspace*{.5ex} \\
\widehat{\Det}(a) & = &
 \left \{
 \begin{array}{l@{\;\;}l}
 \VTrue  & \mathrm{if}\; a \in \set{\VTrue,\VFalse,\VBoth} \\
 \VFalse & \mathrm{otherwise},
 \end{array}
 \right.
\vspace*{.5ex} \\
\widehat{\Confl}(a) & = &
 \left \{
 \begin{array}{l@{\;\;}l}
 \VBoth    & \mathrm{if}\; a = \VNeither \\
 \VNeither & \mathrm{if}\; a = \VBoth    \\
 a         & \mathrm{otherwise}\;.
 \end{array}
 \right.
\end{array}
\]
\end{sdef}

The following is a result concerning the definability of the 
non-classical connectives $\Des$, $\Norm$, $\Cons$, and $\Det$ in
$\BDL^{\IImpl,\False,\Des}$, $\BDL^{\IImpl,\False,\Norm}$, 
$\BDL^{\IImpl,\False,\Cons}$, and $\BDL^{\IImpl,\False,\Det}$, 
respectively.
\begin{theorem}
\label{theorem-definability}
We have: 
\begin{center}
\renewcommand{\arraystretch}{1.25}
\begin{tabular}[t]{l@{\;}l@{\;}l}
$\Des$  & is definable in $\BDL^{\IImpl,\False,\Des}$  & in terms of 
$\set{\Not,\IImpl,\False}$,
\\
$\Norm$ & is definable in $\BDL^{\IImpl,\False,\Norm}$ & in terms of 
$\set{\Not,\CAnd,\COr,\IImpl,\False}$,
\\
$\Cons$ & is definable in $\BDL^{\IImpl,\False,\Cons}$ & in terms of 
$\set{\Not,\CAnd,\IImpl,\False}$,
\\
$\Det$  & is definable in $\BDL^{\IImpl,\False,\Det}$  & in terms of 
$\set{\Not,\COr,\IImpl,\False}$.
\end{tabular}
\end{center}
\end{theorem}
\begin{proof}
From the proof of Theorem~\ref{theorem-lsyn-leqv-bd}, it follows
immediately that Theorem~\ref{theorem-lsyn-leqv-bd} goes through for 
expansions of \BDL.
From this, the following result follows easily:
there exists a $p \in \PVar$ such that
\begin{ldispl}
\renewcommand{\arraystretch}{1.25}
\begin{array}[t]{r@{\;\;}c@{\;\;}l}
\Des p  & \LSyn_{\BDL^{\IImpl,\False,\Des}} & 
\Not (p \IImpl \False)\;,               \\
\Norm p & \LSyn_{\BDL^{\IImpl,\False,\Norm}} & 
((p \CAnd \Not p) \IImpl \False) \CAnd 
\Not ((p \COr \Not p) \IImpl \False)\;, \\
\Cons p & \LSyn_{\BDL^{\IImpl,\False,\Cons}} &
(p \CAnd \Not p) \IImpl \False\;,       \\
\Det p  & \LSyn_{\BDL^{\IImpl,\False,\Det}} & 
\Not ((p \COr \Not p) \IImpl \False)\;.
\end{array}
\end{ldispl}%
From this, the theorem immediately follows.
\qed
\end{proof}
 
The following is a result concerning the non-definability of the 
non-classical connective $\Confl$ in $\BDL^{\IImpl,\False,\Confl}$.
\begin{theorem}
\label{theorem-non-definability}
The connective $\Confl$ is not definable in 
$\BDL^{\IImpl,\False,\Confl}$.
\end{theorem}
\begin{proof}
It is easy to check that, for each unary connective $\Diamond$ with
intended interpretation $\widehat{\Diamond}$ definable in 
$\BDL^{\IImpl,\False,\Diamond}$, 
$\widehat{\Diamond}(\VBoth) \in \set{\VTrue,\VFalse,\VBoth}$ and
$\widehat{\Diamond}(\VNeither) \in \set{\VTrue,\VFalse,\VNeither}$.
Now consider the conflation connective.
Clearly, 
$\widehat{\Confl}(\VBoth) \notin \set{\VTrue,\VFalse,\VBoth}$ and
$\widehat{\Confl}(\VNeither) \notin \set{\VTrue,\VFalse,\VNeither}$.
Hence, $\Confl$ is not definable in $\BDL^{\IImpl,\False,\Confl}$.
\qed 
\end{proof}
As a corollary of Corollary~\ref{corollary-truth-func-compl} and 
Theorem~\ref{theorem-non-definability}, we have that 
$\BDL^{\IImpl,\False}$ is not truth-functionally complete.

\pagebreak[2]

This raises the question which connectives $\Diamond$ are definable in 
$\BDL^{\IImpl,\False,\Diamond}$.
It follows from Theorem~16 in~\cite{AA17a} that an $n$-ary connective
$\Diamond$ with intended interpretation $\widehat{\Diamond}$ is 
definable in $\BDL^{\IImpl,\False,\Diamond}$ iff
\begin{center}
for all $a_1,\ldots,a_n \in \set{\VTrue,\VFalse,\VBoth}$,
$\widehat{\Diamond}(a_1,\ldots,a_n) \in \set{\VTrue,\VFalse,\VBoth}$,
\\
for all $a_1,\ldots,a_n \in \set{\VTrue,\VFalse,\VNeither}$,
$\widehat{\Diamond}(a_1,\ldots,a_n) \in \set{\VTrue,\VFalse,\VNeither}$.
\end{center} 
The connectives $\Des$, $\Norm$, $\Cons$, and $\Det$ belong to a family 
$\indfam
  {\heartsuit_V}{V \subseteq \set{\VTrue,\VFalse,\VBoth,\VNeither}}$
of $16$ non-classical unary connectives, 
where the intended interpretation of the connective $\heartsuit_V$ is 
the function 
\smash{$\funct{\widehat{\heartsuit}_V}
              {\set{\VTrue,\VFalse,\VBoth,\VNeither}}
              {\set{\VTrue,\VFalse,\VBoth,\VNeither}}$}
defined as follows:
\[
\begin{array}[t]{r@{\;\;}c@{\;\;}l}
\widehat{\heartsuit}_V(a) & = &
 \left \{
 \begin{array}{l@{\;\;}l}
 \VTrue  & \mathrm{if}\; a \in V \\
 \VFalse & \mathrm{otherwise}\;.
 \end{array}
 \right.
\end{array}
\]
For each $V \subseteq \set{\VTrue,\VFalse,\VBoth,\VNeither}$, 
$\heartsuit_V$ is definable in $\BDL^{\IImpl,\False,\heartsuit_V}$ 
because, for all $a \in \set{\VTrue,\VFalse,\VBoth}$,
$\widehat{\heartsuit}_V(a) \in \set{\VTrue,\VFalse,\VBoth}$ and,
for all $a \in \set{\VTrue,\VFalse,\VNeither}$,
$\widehat{\heartsuit}_V(a) \in \set{\VTrue,\VFalse,\VNeither}$.
The use of several connectives from this family is not entirely clear.

In addition to the notion of definability from~\cite{SO14a} referred in 
the second last paragraph of Section~\ref{DEFINABILITY}, a weaker notion 
of definability is defined in that paper.
In~\cite{OS15a}, a weaker notion of truth-functional completeness is 
defined in terms of this weaker notion of definability and it is shown 
that BD$\mathrm{\Delta}$ is truth-functionally complete in that weaker 
sense.
From this and the fact that 
$\BDL^{\IImpl,\False} \simeq \mathrm{BD\Delta}$ 
(shown below in Section~\ref{INTERDEFINABILITY-BD}), it follows that 
$\BDL^{\IImpl,\False}$ is truth-functionally complete in that weaker 
sense.

\section{Interdefinability of Expansions of \BDL}
\label{INTERDEFINABILITY-BD}

\sloppy
Recently studied expansions of \BDL\ include the propositional fragments 
of BD$\mathrm{\Delta}$~\cite{SO14a} and QLET$_\mathrm{F}$~\cite{ARCC22a}.
In the notation of this paper, the propositional fragments of these 
expansions are referred to by $\BDL^{\Des}$ and $\BDL^{\Norm}$, 
respectively.
In this section, interdefinability results concerning 
$\BDL^{\IImpl,\False}$, $\BDL^{\Des}$, $\BDL^{\Norm}$, and some other
expansions of \BDL\ are presented.

\begin{theorem}
\label{theorem-implfalse-des}
$\BDL^{\IImpl,\False} \simeq \BDL^{\Des}$.
\end{theorem}
\begin{proof}
It is already known from Theorem~\ref{theorem-definability} that the 
connective $\Des$ from the alphabet of the language of $\BDL^{\Des}$ is 
definable in $\BDL^{\IImpl,\False,\Des}$.
The other way round, the connectives $\IImpl$ and $\False$ from the 
alphabet of the language of $\BDL^{\IImpl,\False}$ are definable 
in $\BDL^{\IImpl,\False,\Des}$ because
\[
\renewcommand{\arraystretch}{1.25}
\begin{array}[t]{r@{\;\;}c@{\;\;}l}
p_1 \IImpl p_2  & \LSyn_{\BDL^{\IImpl,\False,\Des}}
                & \Not (\Des p_1) \COr p_2\;,   \\
\False          & \LSyn_{\BDL^{\IImpl,\False,\Des}}
                & \Des p \CAnd \Not (\Des p)\;,
\end{array}
\]
where $p$ is a fixed but arbitrary propositional variable from $\PVar$.
Hence, \mbox{$\BDL^{\IImpl,\False} \simeq \BDL^{\Des}$}.
\qed
\end{proof}
In other words, $\BDL^{\IImpl,\False}$ is interdefinable with the 
propositional fragment of $\mathrm{BD}\Des$.

\begin{theorem}
\label{theorem-des-consdet}
$\BDL^{\Des} \simeq \BDL^{\Cons,\Det}$ and\,
$\BDL^{\IImpl,\False} \simeq \BDL^{\Cons,\Det}$.
\end{theorem}
\begin{proof}
The connectives $\Cons$ and $\Det$ from the alphabet of the language of 
$\BDL^{\Cons,\Det}$ are definable in $\BDL^{\Des,\Cons,\Det}$ 
because
\[
\renewcommand{\arraystretch}{1.25}
\begin{array}[t]{r@{\;\;}c@{\;\;}l}
\Cons p & \LSyn_{\BDL^{\Des,\Cons,\Det}} 
        & \Not (\Des (p \CAnd \Not p))\;, \\
\Det p  & \LSyn_{\BDL^{\Des,\Cons,\Det}}
        & \Des (p \COr \Not p)\;.
\end{array}
\]
The other way round, the connective $\Des$ from the 
alphabet of the language of $\BDL^{\Des}$ is definable in 
$\BDL^{\Des,\Cons,\Det}$ because
\[
\begin{array}[t]{r@{\;\;}c@{\;\;}l}
\Des p & \LSyn_{\BDL^{\Des,\Cons,\Det}}
       & (p \COr \Not (\Cons p)) \CAnd \Det p\;.
\end{array}
\]
Hence, $\BDL^{\Des} \simeq \BDL^{\Cons,\Det}$.

It is already known from Theorem~\ref{theorem-implfalse-des} that
moreover $\BDL^{\IImpl,\False} \simeq \BDL^{\Des}$.
Because $\simeq$ is an equivalence relation, it follows from 
$\BDL^{\IImpl,\False} \simeq \BDL^{\Des}$ and
$\BDL^{\Des} \simeq \BDL^{\Cons,\Det}$ that also
$\BDL^{\IImpl,\False} \simeq \BDL^{\Cons,\Det}$.
\qed
\end{proof}

\begin{theorem}
\label{theorem-consdet-norm}
$\BDL^{\Norm}$ is definable in $\BDL^{\Cons,\Det}$,
$\BDL^{\Norm} \not\simeq \BDL^{\Cons,\Det}$, and\,
$\BDL^{\IImpl,\False} \not\simeq \BDL^{\Norm}$.
\end{theorem}
\begin{proof}
The connective $\Norm$ from the alphabet of the language of 
$\BDL^{\Norm}$ is definable in $\BDL^{\Cons,\Det,\Norm}$ because
\[
\begin{array}[t]{r@{\;\;}c@{\;\;}l}
\Norm p & \LSyn_{\BDL^{\Cons,\Det,\Norm}} & \Cons p \CAnd \Det p\;.
\end{array}
\]
Hence, $\BDL^{\Norm}$ is definable in $\BDL^{\Cons,\Det}$.

It is easy to check that, for each unary connective $\Diamond$ with
intended interpretation $\widehat{\Diamond}$ definable in 
$\BDL^{\Norm,\Diamond}$, either
$\widehat{\Diamond}(\VBoth) = \VBoth$ and
$\widehat{\Diamond}(\VNeither) = \VNeither$ or
$\widehat{\Diamond}(\VBoth) = \widehat{\Diamond}(\VNeither)
 \in \set{\VTrue,\VFalse}$.
Clearly, $\widehat{\Cons}(\VBoth) = \VFalse$, 
$\widehat{\Cons}(\VNeither) = \VTrue$, 
$\widehat{\Det}(\VBoth) = \VTrue$, and
$\widehat{\Det}(\VNeither) = \VFalse$.
This means that the connectives $\Cons$ and $\Det$ from the alphabet of 
the language of $\BDL^{\Cons,\Det}$ are not definable in 
$\BDL^{\Cons,\Det,\Norm}$.
Hence, $\BDL^{\Norm} \not\simeq \BDL^{\Cons,\Det}$.

Because it is known from Theorem~\ref{theorem-des-consdet} that
$\BDL^{\IImpl,\False} \simeq \BDL^{\Cons,\Det}$, we also have 
$\BDL^{\IImpl,\False} \not\simeq \BDL^{\Norm}$.
\qed
\end{proof}
In other words, $\BDL^{\IImpl,\False}$ is not interdefinable with the 
propositional fragment of $\mathrm{QLET_F}$.
However, because $\BDL^{\Norm}$ is definable in $\BDL^{\Cons,\Det}$ and
$\BDL^{\IImpl,\False} \simeq \BDL^{\Cons,\Det}$, the propositional 
fragment of $\mathrm{QLET_F}$ is definable in $\BDL^{\IImpl,\False}$.

The following is a corollary of Theorem~\ref{theorem-non-definability}
and its proof.
\begin{corollary}
$\BDL^{\IImpl,\False} \not\simeq \BDL^{\Confl}$ and\,
$\BDL^{\IImpl,\False} \not\simeq \BDL^{\IImpl,\Confl}$.
\end{corollary}
We know from~\cite{AA17a} (Theorems~11 and~16) that both
$\BDL^{\IImpl,\False}$ and $\BDL^{\IImpl,\Confl}$ are 
not truth-functionally complete.
We also know from~\cite{AA17a} (Theorem~4) that 
$\BDL^{\IImpl,\Both,\Neither}$, where $\Both$ and $\Neither$ are the 
nullary connectives with the intended interpretations 
\smash{$\widehat{\Both} = \VBoth$} and 
\smash{$\widehat{\Neither} = \VNeither$}, 
is truth-functionally complete.
We have the following result concerning $\BDL^{\IImpl,\Both,\Neither}$.
\begin{theorem}
\label{theorem-FvsBandN}
$\BDL^{\IImpl,\False} \not\simeq \BDL^{\IImpl,\Both,\Neither}$, but
$\BDL^{\IImpl,\False}$ is definable in $\BDL^{\IImpl,\Both,\Neither}$.
\end{theorem}
\begin{proof}
We know that $\BDL^{\IImpl,\False}$ is not truth-functionally complete 
and $\BDL^{\IImpl,\Both,\Neither}$ is truth-functionally complete.
From this and Corollary~\ref{corollary-truth-func-compl}, the theorem 
follows immediately.
\qed
\end{proof}
It follows immediately from Theorem~\ref{theorem-FvsBandN} that
the connective $\False$ from the alphabet of the language of 
$\BDL^{\IImpl,\False}$ is definable in 
$\BDL^{\IImpl,\Both,\Neither,\False}$.
This also follows immediately from the easily established synonymity
\[
\begin{array}[t]{r@{\;\;}c@{\;\;}l}
\False & \LSyn_{\BDL^{\IImpl,\Both,\Neither,\False}}
       & \Both \CAnd \Neither\;.
\end{array}
\]

\section{$\BDL^{\IImpl,\False}$ and Classical Logic}
\label{CLASSICAL-LOGIC}

Roughly speaking, a connective of a non-classical logic is considered 
a classical connective if it has many properties in common with a 
connective known from classical logic, in particular properties in 
relation to logical consequence.
This section addresses the question to what extent the connectives of 
$\BDL^{\IImpl,\False}$ are classical connectives.

The first part of the next proposition concerns properties of the 
connectives $\CAnd$, $\COr$, and $\IImpl$ in relation to the logical 
consequence relation of $\BDL^{\IImpl,\False}$ and the second part of 
the next proposition concerns an indirect property of the connective 
$\Not$ in relation to the logical consequence relation of 
$\BDL^{\IImpl,\False}$.
Both parts follow easily from the definition of this logical consequence 
relation.
\begin{proposition}
\label{prop-normal}
\mbox{}
\begin{enumerate}
\item[\textup{1}.]
$\BDL^{\IImpl,\False}$ is \emph{normal}, i.e.\
$\LCon$ is such that for all $\Gamma, \Delta \subseteq \LForm$, and
$A_1, A_2 \in \LForm$:
\[
\renewcommand{\arraystretch}{1.25}
\begin{array}[t]{r@{\;}c@{\;}l@{\;\;}l}
\Gamma \LCon \Delta, A_1 \CAnd A_2    & \mathrm{iff} &
\multicolumn{2}{l}
{\Gamma \LCon \Delta, A_1 \;\mathrm{and}\;
 \Gamma \LCon \Delta, A_2\;,}
\\
A_1 \COr A_2, \Gamma \LCon \Delta     & \mathrm{iff} &
\multicolumn{2}{l}
{A_1, \Gamma \LCon \Delta \;\mathrm{and}\;
 A_2, \Gamma \LCon \Delta\;,}
\\
\Gamma \LCon \Delta, A_1 \IImpl A_2   & \mathrm{iff} & 
A_1, \Gamma \LCon \Delta, A_2\;;
\end{array}
\]
\item[\textup{2}.]
$\BDL^{\IImpl,\False}$ is $\Not$-\emph{contained in classical logic}, 
i.e.\ 
there exists a logic with the same language as $\BDL^{\IImpl,\False}$ 
and a logical consequence relation $\LCon'$ such that:
\begin{itemize}
\item
${\LCon} \subseteq {\LCon'}$;
\item
$\LCon'$ is induced by a matrix 
$\langle \TValue', \DValue', \TFunct' \rangle$  such that 
$\TValue' = \set{\VTrue,\VFalse}$, 
$\DValue' = \set{\VTrue}$, and $\TFunct'(\Not)$ is defined as follows:
\[
\renewcommand{\arraystretch}{1.25}
\begin{array}[t]{c}
\TFunct'(\Not)(a) =
 \left \{
 \begin{array}{l@{\;\;}l}
 \VTrue  & \mathrm{if}\; a = \VFalse \\
 \VFalse & \mathrm{if}\; a = \VTrue\;,
 \end{array}
 \right.
\end{array}
\]
where $a$ ranges over all truth values in $\TValue'$.
\end{itemize}
\end{enumerate}
\end{proposition}
Clearly, $\BDL^{\IImpl,\False}$ shares these properties with 
$\CL^{\IImpl,\False}$.

The following two properties of the connective $\Not$ in relation to the 
logical consequence relation of $\BDL^{\IImpl,\False}$ also follow 
easily from the definition of this logical consequence relation.
\begin{proposition}
\mbox{}
\begin{enumerate}
\item[\textup{1}.]
there exist a $\Gamma \subseteq \LForm$ and $A, A' \in \LForm$ such that 
$\Gamma \LCon A$ and $\Gamma \LCon \Not A$, but $\Gamma \not\LCon A'$;
\item[\textup{2}.]
there exist a $\Gamma \subseteq \LForm$ and $A, A' \in \LForm$ such that 
$\Gamma, A \LCon A'$ and $\Gamma, \Not A \LCon A'$, but 
$\Gamma \not\LCon A'$.
\end{enumerate}
\end{proposition}
Clearly, $\BDL^{\IImpl,\False}$ does not share these properties with 
$\CL^{\IImpl,\False}$.
Because $\BDL^{\IImpl,\False}$ is normal and $\Not$-contained in 
classical logic, these properties imply that $\BDL^{\IImpl,\False}$ is 
\emph{paraconsistent} and \emph{paracomplete}, respectively, in the 
sense of~\cite{AA17a}.

One way to illustrate how similar the connectives of 
$\BDL^{\IImpl,\False}$ and $\CL^{\IImpl,\False}$ are with respect to 
logical consequence is to show how sound and complete sequent calculus 
proof systems for these logics are related to each other.

A sequent calculus proof system for $\BDL^{\IImpl,\False}$ is given in 
Table~\ref{table-proof-system}.
\begin{table}[!t]
\caption{A sequent calculus proof system for $\BDL^{\IImpl,\False}$}
\label{table-proof-system}
\vspace*{-2ex} 
\renewcommand{\arraystretch}{.5} 
\centering
\begin{tabular}[t]{@{}c@{}}
\hline
\\[-1ex]
\begin{small}
\begin{tabular}{@{}l@{}}
\InfRule{Id}
 {\phantom{\Gamma \LEnt \Delta}}
 {A, \Gamma \LEnt \Delta, A}
\\[3ex] 
\InfRule{$\False$-L}
 {{}}
 {\False, \Gamma \LEnt \Delta}
\\[3ex] 
\InfRule{$\CAnd$-L}
 {A\sb1, A\sb2, \Gamma \LEnt \Delta}
 {A\sb1 \CAnd A\sb2, \Gamma \LEnt \Delta}
\\[3ex]
\InfRule{$\COr$-L}
 {A\sb1, \Gamma \LEnt \Delta \quad
  A\sb2, \Gamma \LEnt \Delta}
 {A\sb1 \COr A\sb2, \Gamma \LEnt \Delta}
\\[3ex]
\InfRule{$\IImpl$-L}
 {\Gamma \LEnt \Delta, A\sb1 \quad
  A\sb2, \Gamma \LEnt \Delta}
 {A\sb1 \IImpl A\sb2, \Gamma \LEnt \Delta}
\\[3ex]
{}
\\[3ex] 
\InfRule{$\Not \Not$-L}
 {A, \Gamma \LEnt \Delta}
 {\Not \Not A, \Gamma \LEnt \Delta}
\\[3ex]
\InfRule{$\Not \CAnd$-L}
 {\Not A\sb1, \Gamma \LEnt \Delta \quad
  \Not A\sb2, \Gamma \LEnt \Delta}
 {\Not (A\sb1 \CAnd A\sb2), \Gamma \LEnt \Delta}
\\[3ex]
\InfRule{$\Not \COr$-L}
 {\Not A\sb1, \Not A\sb2, \Gamma \LEnt \Delta}
 {\Not (A\sb1 \COr A\sb2), \Gamma \LEnt \Delta}
\\[3ex]
\InfRule{$\Not \IImpl$-L}
 {A\sb1, \Not A\sb2, \Gamma \LEnt \Delta}
 {\Not (A\sb1 \IImpl A\sb2), \Gamma \LEnt \Delta}
\\[3ex]
\end{tabular}
\qquad \qquad
\begin{tabular}{@{}l@{}} 
\InfRule{Cut}
 {\Gamma \LEnt \Delta, A \quad
  A, \Gamma' \LEnt \Delta'}
 {\Gamma', \Gamma \LEnt \Delta, \Delta'}
\\[3ex] 
{}
\\[3ex] 
\InfRule{$\CAnd$-R}
 {\Gamma \LEnt \Delta, A\sb1 \quad
  \Gamma \LEnt \Delta, A\sb2}
 {\Gamma \LEnt \Delta, A\sb1 \CAnd A\sb2}
\\[3ex] 
\InfRule{$\COr$-R}
 {\Gamma \LEnt \Delta, A\sb1, A\sb2}
 {\Gamma \LEnt \Delta, A\sb1 \COr A\sb2}
\\[3ex]
\InfRule{$\IImpl$-R}
 {A\sb1, \Gamma \LEnt \Delta, A\sb2}
 {\Gamma \LEnt \Delta, A\sb1 \IImpl A\sb2}
\\[3ex]
\InfRule{$\Not \False$-R}
 {{}}
 {\Gamma \LEnt \Delta, \Not \False}
\\[3ex] 
\InfRule{$\Not \Not$-R}
 {\Gamma \LEnt \Delta, A}
 {\Gamma \LEnt \Delta, \Not \Not A}
\\[3ex] 
\InfRule{$\Not \CAnd$-R}
 {\Gamma \LEnt \Delta, \Not A\sb1, \Not A\sb2}
 {\Gamma \LEnt \Delta, \Not (A\sb1 \CAnd A\sb2)}
\\[3ex] 
\InfRule{$\Not \COr$-R}
 {\Gamma \LEnt \Delta, \Not A\sb1 \quad
  \Gamma \LEnt \Delta, \Not A\sb2}
 {\Gamma \LEnt \Delta, \Not (A\sb1 \COr A\sb2)}
\\[3ex] 
\InfRule{$\Not \IImpl$-R}
 {\Gamma \LEnt \Delta, A\sb1 \quad
  \Gamma \LEnt \Delta, \Not A\sb2}
 {\Gamma \LEnt \Delta, \Not (A\sb1 \IImpl A\sb2)}
\\[3ex]
\end{tabular}
\end{small}
\\[-1ex]
\hline
\end{tabular}
\end{table}
In this table, 
$A,A_1,\mathrm{and}\;A_2$ are meta-variables ranging over all formulas 
from $\LForm$ and
$\Gamma$, $\Gamma'$, $\Delta$, and $\Delta'$ are meta-variables ranging 
over all finite sets of formulas from $\LForm$. 
The sequent calculus proof system of $\BDL^{\IImpl,\False}$ is sound 
and complete with respect to the logical consequence relation of
$\BDL^{\IImpl,\False}$, i.e., for all 
$\Gamma,\Delta \subseteq \LForm$, $\Gamma \LCon \Delta$ iff
there exist finite sets $\Gamma' \subseteq \Gamma$ and 
$\Delta' \subseteq \Delta$ such that $\Gamma' \LEnt \Delta'$ is 
provable (Theorem~20 from~\cite{AA17a}).
A sound and complete sequent calculus proof system of 
$\CL^{\IImpl,\False}$ can be obtained by adding the following two 
inference rules to the sequent calculus proof system of 
$\BDL^{\IImpl,\False}$: 
\begin{center}
\begin{tabular}{@{}l@{}}
\InfRule{$\Not$-L}
 {\Gamma \LEnt \Delta, A}
 {\Not A, \Gamma \LEnt \Delta}
\qquad\qquad
\InfRule{$\Not$-R}
 {A, \Gamma \LEnt \Delta}
 {\Gamma \LEnt \Delta, \Not A}
\end{tabular}
\end{center}
By the addition of these inference rules, the rules from 
Table~\ref{table-proof-system} whose name begins with $\Not$ become
derived inference rules.

$\BDL^{\IImpl,\False}$ is not the only logic that is normal and
$\Not$-contained in classical logic.
Any logic with the same language as $\BDL^{\IImpl,\False}$ and a logical
consequence relation  that is induced by a strongly regular four-valued
matrix is normal and $\Not$-contained in classical logic.
\begin{sdef}
\label{def-strongly-regular}
\sloppy
Let $\Matrix = (\TValue,\DValue,\TFunct)$ be a four-valued matrix.
Then $\Matrix$ is a \emph{strongly regular four-valued matrix} iff:
\begin{itemize}
\item
$\Matrix$ is a matrix for the alphabet of the language of 
$\BDL^{\IImpl,\False}$;
\item
$\TValue = \set{\VTrue,\VFalse,\VBoth,\VNeither}$ and  
$\DValue = \set{\VTrue,\VBoth}$;
\item
$
\renewcommand{\arraystretch}{1.25}
\begin{array}[t]{l@{\;}c@{\;}l} 
\TFunct(\False) = \VFalse\;,           
\\
\TFunct(\Not)(a) \in \DValue         & \mathrm{iff} & 
a \in \set{\VFalse,\VBoth}\;,
\\
\TFunct(\CAnd)(a_1,a_2) \in \DValue  & \mathrm{iff} &
a_1 \in \DValue \;\mathrm{and}\; a_2 \in \DValue\;, 
\\
\TFunct(\COr)(a_1,a_2) \in \DValue   & \mathrm{iff} &
a_1 \in \DValue \;\mathrm{or}\; a_2 \in \DValue\;, 
\\
\TFunct(\IImpl)(a_1,a_2) \in \DValue & \mathrm{iff} &
a_1 \in \NDValue \;\mathrm{or}\; a_2 \in \DValue\;; 
\end{array}
$
\item 
for all $a_1,a_2 \in \set{\VTrue,\VFalse}$,
$\TFunct(\Not)(a_1) \in \set{\VTrue,\VFalse}$,
$\TFunct(\CAnd)(a_1,a_2) \in \set{\VTrue,\VFalse}$,
$\TFunct(\COr)(a_1,a_2) \in \set{\VTrue,\VFalse}$, and
$\TFunct(\IImpl)(a_1,a_2) \in \set{\VTrue,\VFalse}$.
\end{itemize}
\end{sdef}

The following result is a corollary of Definitions~\ref{def-Matrix-BD} 
and~\ref{def-strongly-regular}.
\begin{corollary}
\label{corollary-strongly-regular}
The four-valued simple matrix that induces the logical consequence 
relation of $\BDL^{\IImpl,\False}$ is a strongly regular four-valued 
matrix.
\end{corollary}

All strongly regular four-valued matrices seem quite similar.
Let $\Matrix = (\TValue,\DValue,\TFunct)$ be a strongly regular 
four-valued matrix.
Then 
\[
{\LCon_{\CLp^{\IImpl,\False}}} \subset {\LCon_\Matrix} \subset
{\LCon_{\CL^{\IImpl,\False}}}\;.
\]
Moreover, whereas the conditions imposed on $\TFunct(\CAnd)$, 
$\TFunct(\COr)$, and $\TFunct(\IImpl)$ agree exactly with the intuition 
that $\VBoth$ is just an alternative for $\VTrue$ and $\VNeither$ is 
just an alternative for $\VFalse$, the conditions imposed on 
$\TFunct(\Not)$ agree exactly with the intuition that $\VTrue$ conveys 
that its negation is $\VFalse$, $\VFalse$ conveys that its negation is 
$\VTrue$, $\VBoth$ conveys that its negation is a designated truth 
value, and $\VNeither$ conveys that its negation is not a designated 
truth value.
Thus, the differences between the logical consequence relations that 
different strongly regular four-valued matrices induce are all due to 
small differences in the interpretation of $\Not$.
However, under all allowed interpretations ($p \in \PVar$):
\[
p \not\LCon_\Matrix \Not p
\quad \mathrm{and}\quad 
\Not p \not\LCon_\Matrix p\;.
\]

Although all strongly regular four-valued matrices seem quite similar, 
the simple matrix that induces the logical consequence relation of 
$\BDL^{\IImpl,\False}$ is just one of the $2^{38}$ strongly regular 
four-valued matrices.
So, $\BDL^{\IImpl,\False}$ is one of at least $2^{38}$ logics that are
normal and $\Not$-contained in classical logic.
The question is whether the connectives of $\BDL^{\IImpl,\False}$ are 
more classical than those of the other logics induced by a strongly 
regular four-valued matrix.
Some insight into this issue is acquired by considering to what extent 
the connectives of $\BDL^{\IImpl,\False}$ and the other logics induced 
by a strongly regular four-valued matrix are similar in terms of logical 
equivalence.

Among the logics induced by a strongly regular four-valued matrix, 
$\BDL^{\IImpl,\False}$ is the only one with a logical equivalence 
relation that satisfies all laws given in Table~\ref{laws-lequiv} 
(this result is a corollary of the proof of the corresponding theorem 
for a first-order version of $\BDL^{\IImpl,\False}$ from~\cite{Mid23a}).
\begin{table}[!t]
\caption{The distinguishing laws of logical equivalence for
 $\BDL^{\IImpl,\False}$}
\label{laws-lequiv}
\begin{eqntbl}
\renewcommand{\arraystretch}{1.2}
\begin{eqncol}
(1)  & A \CAnd \False \LEqv \False \\
(3)  & A \CAnd \True \LEqv A \\
(5)  & A \CAnd A \LEqv A \\
(7)  & A_1 \CAnd A_2 \LEqv A_2 \CAnd A_1 \\
(9)  & \Not (A_1 \CAnd A_2) \LEqv \Not A_1 \COr \Not A_2 \\
(11)  & \Not \Not A \LEqv A \\
(12)  & (A_1 \CAnd (A_1 \IImpl \False)) \IImpl A_2 \LEqv \True 
\end{eqncol}
\qquad
\begin{eqncol}
(2)  & A \COr \True \LEqv \True \\
(4)  & A \COr \False \LEqv A \\
(6)  & A \COr A \LEqv A \\
(8)  & A_1 \COr A_2 \LEqv A_2 \COr A_1 \\
(10)  & \Not (A_1 \COr A_2) \LEqv \Not A_1 \CAnd \Not A_2 \\
\\
(13)  & (A_1 \COr (A_1 \IImpl \False)) \IImpl A_2 \LEqv A_2  
\end{eqncol}
\end{eqntbl}
\end{table}
Laws (1)--(11) are basic classical laws of logical equivalence.
$\BDL^{\IImpl,\False}$ is one of $2^{12}$ logics induced by a strongly 
regular four-valued matrix with a logical equivalence relation that 
satisfies laws~(1)--(11).
Laws (12) and (13) follow from the other laws and the following 
classical law of logical equivalence: 
$\Not (A_1 \IImpl A_2) \LEqv A_1 \CAnd \Not A_2$.
However, this law is not satisfied by the logical equivalence relation 
of $\BDL^{\IImpl,\False}$.%
\footnote
{Consequently, in the case of $\BDL^{\IImpl,\False}$, not every formula
is logically equivalent to a formula in conjunctive normal form.}

Put in other words, with respect to logical equivalence, the connectives 
$\Not$, $\CAnd$, and $\COr$ of $\BDL^{\IImpl,\False}$ are more classical 
than the connectives $\Not$, $\CAnd$, and $\COr$ of the other logics 
induced by a strongly regular four-valued matrix, but it is not clear 
whether, with respect to logical equivalence, the connective $\IImpl$ of 
$\BDL^{\IImpl,\False}$ is more classical than the connective $\IImpl$ of 
the other logics induced by a strongly regular four-valued matrix.
Moreover, the question remains whether properties of connectives with 
respect to logical equivalence should be considered relevant to the 
extent to which they are classical.
One reason to consider them relevant is that, unlike in classical logic, 
it is not the case in every logic that logical equivalence can be 
defined solely in terms of logical consequence.

\section{More on $\BDL^{\IImpl,\False}$ and Classical Logic}
\label{MORE-CLASSICAL-LOGIC}

In previous sections, it has been shown that the connectives of 
$\BDL^{\IImpl,\False}$ have many properties with regard to logical 
consequence in common with connectives known from classical logic.
Due to this classical nature of the connectives of 
$\BDL^{\IImpl,\False}$, there exists a simple translation of the 
formulas of $\BDL^{\IImpl,\False}$ to formulas of $\CL^{\IImpl,\False}$ 
that preserves logical consequence.
This means that the translation in question provides an embedding 
of $\BDL^{\IImpl,\False}$ into $\CL^{\IImpl,\False}$.
 
The existence of such a translation is practically relevant.
To give an example, the translation can be useful to determine, 
for a fragment for which logical consequence is decidable in 
$\CL^{\IImpl,\False}$, whether logical consequence is decidable in 
$\BDL^{\IImpl,\False}$ and to adapt, for such a fragment, an existing 
decision procedure for logical consequence in $\CL^{\IImpl,\False}$ 
to logical consequence in $\BDL^{\IImpl,\False}$.
That is why some attention is paid to the translation in question in 
this section. 

The translation is given by a function from the set of all formulas of a 
fixed but arbitrary instance of $\BDL^{\IImpl,\False}$ to the set of all 
formulas of an instance of $\CL^{\IImpl,\False}$.
Let $\PVar$ be the set of all propositional variables of this instance of 
$\BDL^{\IImpl,\False}$.
Then the set $\PVar'$ of all propositional variables of the instance of 
$\CL^{\IImpl,\False}$ concerned is $\PVar$ extended as follows: 
$\PVar' = \PVar \union \set{\denial{p} \where p \in \PVar}$.

The translation function, denoted by $\Embed{\ph}{}{}$, is inductively 
defined in Table~\ref{table-translation}. 
\begin{table}[!t]
\caption{Translation of the formulas of $\BDL^{\IImpl,\False}$}
\label{table-translation}
\vspace*{-3ex} \par \mbox{} \centering
\renewcommand{\arraystretch}{1.275}
\begin{array}[t]{rcl}
\hline
\mbox{} \\[-2.5ex]
\Embed{p}{}{} & = & p
\\
\Embed{\False}{}{} & = & \False 
\\
\Embed{A\sb1 \CAnd A\sb2}{}{} & = &
  \Embed{A\sb1}{}{} \CAnd \Embed{A\sb2}{}{} 
\\
\Embed{A\sb1 \COr A\sb2}{}{} & = &
  \Embed{A\sb1}{}{} \COr \Embed{A\sb2}{}{} 
\\
\Embed{A\sb1 \IImpl A\sb2}{}{} & = &
  \Embed{A\sb1}{}{} \IImpl \Embed{A\sb2}{}{} 
\\[1ex]
\Embed{\Not p}{}{} & = & \denial{p}
\\
\Embed{\Not \False}{}{} & = & \Not \False 
\\
\Embed{\Not \Not A}{}{} & = & \Embed{A}{}{} 
\\
\Embed{\Not (A\sb1 \CAnd A\sb2)}{}{} & = &
  \Embed{\Not A\sb1 \COr \Not A\sb2}{}{} 
\\
\Embed{\Not (A\sb1 \COr A\sb2)}{}{} & = &
  \Embed{\Not A\sb1 \CAnd \Not A\sb2}{}{}
\\
\Embed{\Not (A\sb1 \IImpl A\sb2)}{}{} & = &
  \Embed{A\sb1 \CAnd \Not A\sb2}{}{} 
\vspace*{1.25ex} \\
\hline
\end{array}
\vspace*{-1.5ex} \par
\end{table}
In this table, 
$p$ is a syntactic variable ranging over all propositional variables of 
the fixed instance of $\BDL^{\IImpl,\False}$, and
$A_1$, $A_2$, and $A$ are syntactic variables ranging over all formulas 
of the fixed instance of $\BDL^{\IImpl,\False}$.

The intuition is that $\Embed{A}{}{}$ is a classical-logic formula
stating that the formula $A$ is either true only or both true and false 
in $\BDL^{\IImpl,\False}$.

The given translation provides a simple embedding of 
$\BDL^{\IImpl,\False}$ into $\CL^{\IImpl,\False}$.
\begin{theorem}
\label{theorem-embed}
Let $\LCon_{\BDL^{\IImpl,\False}}$ and $\LCon_{\CL^{\IImpl,\False}}$ be
the logical consequence relations of $\BDL^{\IImpl,\False}$ and
$\CL^{\IImpl,\False}$, respectively. 
Then:
\[
\Gamma \LCon_{\BDL^{\IImpl,\False}} \Delta
\quad \mathrm{iff} \quad
\set{\Embed{A'}{}{} \where A' \in \Gamma} \LCon_{\CL^{\IImpl,\False}}
\set{\Embed{A'}{}{} \where A' \in \Delta}\;.
\]
\end{theorem}
\begin{proof}
This is proved in~\cite{BCK99a} for CLoNs, the variant with a 
single-conclusion logical consequence relation of the expansion of 
$\BDL^{\IImpl,\False}$ with a bi-implication connective (see Theorem~1 
and the remark in the second paragraph on page~42 in that paper). 
The proof for $\BDL^{\IImpl,\False}$ goes similarly because it depends 
neither on the additional connective of CLoNs nor on the restriction to
a single-conclusion logical consequence relation.
\qed
\end{proof}

In the following remark we use the term ``weak negation normal form'' 
several times.
A formula $A$ of $\BDL^{\IImpl,\False}$ is in 
\emph{weak negation normal form} if each occurrences of the connective 
$\Not$ in $A$ is in a subformula of the form $\Not A'$ 
where $A'$ is an atomic formula.

Theorem~\ref{theorem-embed} shows indirectly how close 
$\BDL^{\IImpl,\False}$ and $\CL^{\IImpl,\False}$ are to each other.
Every formula of $\BDL^{\IImpl,\False}$, like every formula of 
$\CL^{\IImpl,\False}$, has a weak negation normal form.
For a formula in weak negation normal form, the given translation causes 
only minor changes. 
It consists solely of replacing each subformula of the form $\Not p$, 
where $p$ is a propositional variable, by $\denial{p}$.
Since a weak negation normal form can be obtained in polynomial time, 
Theorem~\ref{theorem-embed} also shows indirectly that logical 
consequence in $\BDL^{\IImpl,\False}$ is polynomially reducible to 
logical consequence in $\CL^{\IImpl,\False}$.

Not only can $\BDL^{\IImpl,\False}$ be embedded into 
$\CL^{\IImpl,\False}$, $\CL^{\IImpl,\False}$ can be recaptured in
$\BDL^{\IImpl,\False}$.
This is made precise in the following theorem, in which we write
$\mathrm{Cl}(\Gamma)$, where $\Gamma \subseteq \LForm$, for the set
$\set{((p \CAnd \Not p) \IImpl \False) \CAnd
      \Not ((p \COr \Not p) \IImpl \False) \where
      p \in \mathrm{var}(\Gamma)}$.
\begin{theorem}
\label{recapture}
Let $\LCon_{\CL^{\IImpl,\False}}$ and $\LCon_{\BDL^{\IImpl,\False}}$ be
the logical consequence relations of $\CL^{\IImpl,\False}$ and 
$\BDL^{\IImpl,\False}$, respectively. 
Then:
\[
\Gamma \LCon_{\CL^{\IImpl,\False}} \Delta
\quad \mathrm{iff} \quad
\Gamma, \mathrm{Cl}(\Gamma \union \Delta)
 \LCon_{\BDL^{\IImpl,\False}} \Delta\;.
\]
\end{theorem}
\begin{proof}
Let $\Matrix$ be the four-valued simple matrix that induces the logical 
consequence relation of $\BDL^{\IImpl,\False}$ 
and $\Matrix'$ be the two-valued simple matrix that induces the logical 
consequence relation of $\CL^{\IImpl,\False}$.
Moreover, let $\nu$ be a valuation in $\Matrix$.
Then, by Definition~\ref{def-nonclass-conn} and the proof of 
Theorem~\ref{theorem-definability}, 
for all $p \in \mathrm{var}(\Gamma \union \Delta)$,
$\LVal{((p \CAnd \Not p) \IImpl \False) \CAnd
      \Not ((p \COr \Not p) \IImpl \False)} \in \DValue$ iff 
$\LVal{p} \in \set{\VTrue,\VFalse}$.
From this it easily follows, 
using Corollary~\ref{corollary-strongly-regular}, that, 
for all valuations $\nu'$ in $\Matrix'$ that agree with $\nu$ on 
$\mathrm{var}(\Gamma \union \Delta)$,
for all $A \in \Gamma \union \Delta$, $\LVal{A} = \nu'(A)$.
From this, the theorem follows immediately.
\qed
\end{proof}

\section{Concluding Remarks}
\label{CONCLUSIONS}

We have gained some insight into the interdefinability of several 
expansions of Belnap-Dunn logic by investigating the question whether 
the expansions whose connectives include one or more non-classical 
connectives are interdefinable with an expansion whose connectives 
include only classical connectives:
\begin{itemize}
\item
$\BDL^{\Des}$ and $\BDL^{\Cons,\Det}$ are interdefinable with 
$\BDL^{\IImpl,\False}$; 
\item
$\BDL^{\Norm}$ and $\BDL^{\Confl}$ are not interdefinable with 
$\BDL^{\IImpl,\False}$;
\item
$\BDL^{\Norm}$ is definable in $\BDL^{\IImpl,\False}$;
\item
$\BDL^{\IImpl,\False}$ is definable in $\BDL^{\Confl}$.
\end{itemize}
This means, among other things, that $\BDL^{\Des}$, $\BDL^{\Cons,\Det}$,
and $\BDL^{\IImpl,\False}$ can be replaced by each other, that 
$\BDL^{\Norm}$ can be replaced by $\BDL^{\IImpl,\False}$, but that
$\BDL^{\Confl}$ cannot be replaced by $\BDL^{\IImpl,\False}$.

The main advantages of choosing $\BDL^{\IImpl,\False}$ over expansions 
of \BDL\ that are definable in $\BDL^{\IImpl,\False}$ are:
\begin{itemize}
\item
the language of $\BDL^{\IImpl,\False}$ is the same as the language of a 
common version of classical logic;
\item
the logical consequence relation of $\BDL^{\IImpl,\False}$ is included 
in the logical consequence relation of that version of classical logic;
\item
there exists a simple translation of the formulas of 
$\BDL^{\IImpl,\False}$ to formulas of $\CL^{\IImpl,\False}$ that 
preserves logical consequence.
\end{itemize}
The third advantage mentioned above is further discussed in 
Section~\ref{MORE-CLASSICAL-LOGIC}.
The other two advantages mentioned above entail that each of the axioms 
and inference rules of a proof system for $\BDL^{\IImpl,\False}$ is a 
classical one or can be derived from classical ones (see, e.g., the 
sequent calculus proof system for $\BDL^{\IImpl,\False}$ presented in 
Section~\ref{MORE-CLASSICAL-LOGIC}).
Moreover, the missing classical axioms and inference rules are all 
related to the fact that $A, \Not A \LCon \False$ 
(law of non-contradiction) and $\True \LCon A, \Not A$ 
(law of excluded middle) do not hold for all formulas $A$ in the case of 
$\BDL^{\IImpl,\False}$.
Together this means that proving something in $\BDL^{\IImpl,\False}$ 
goes pretty much the same as proving something in $\CL^{\IImpl,\False}$.

In $\BDL^{\IImpl,\False}$, as in classical logic, logically equivalent 
formulas can always be replaced by each other.
Therefore, an additional advantage of choosing $\BDL^{\IImpl,\False}$ is 
that the logical equivalence relation of $\BDL^{\IImpl,\False}$ is 
included in the logical equivalence relation of the version of classical 
logic with the same language.
This entails that each of the laws of logical equivalence that hold for 
the logical equivalence relation of $\BDL^{\IImpl,\False}$ is a 
classical law of logical equivalence.
This means that reasoning about logical equivalence of formulas in 
$\BDL^{\IImpl,\False}$ goes pretty much the same as reasoning about 
logical equivalence of formulas in $\CL^{\IImpl,\False}$.

The fact that $\BDL^{\Confl}$ is not definable in $\BDL^{\IImpl,\False}$
raises the question of what is missing in practice in an expansion of 
\BDL\ whose connectives do not include the conflation connective 
$\Confl$.
This question is difficult to answer.
The problem with the conflation connective is that it has no obvious 
intuitive meaning and there appear to be no practical examples of its 
use.
Theoretically interesting, however, is that the combination of $\Confl$ 
and $\Not$ corresponds to classical negation and, consequently, 
$A, \Confl \Not A \LCon \False$ and $\True \LCon A, \Confl \Not A$ hold 
for all formulas $A$ in the case of $\BDL^{\Confl}$.

In~\cite{Mid23b}, a first-order version of $\BDL^{\IImpl,\False}$ is
presented and applied in the area of relational database theory.
The results concerning the interdefinability of $\BDL^{\IImpl,\False}$
with other expansions of \BDL\ presented in this paper carry over to
the first-order~case.

\bibliographystyle{splncs04}
\bibliography{PCL}

\end{document}